\documentclass[preprint,showpacs,preprintnumbers,amssymb,nofootinbib]{revtex4}
\newcommand{\fslash}[1]{\ooalign{\hfil/\hfil\crcr$#1$}}

\begin{document}
\title{In-medium pion weak decay constants }
\author{Hungchong Kim}
\email{hung@phya.yonsei.ac.kr}
\affiliation{%
Institute of Physics and Applied Physics, Yonsei University,
Seoul 120-749, Korea
}
\begin{abstract}

In nuclear matter, the pion weak decay constant is separated into the two
components $f_t, f_s$ corresponding to the time and space components of 
the axial-vector current.  Using QCD sum rules, we compute the two decay 
constants from the pseudoscalar-axial vector correlation function in the 
matter $i \int d^4x~ e^{ip\cdot x} \langle \rho| T[{\bar d}(x) i 
\gamma_5 u (x)~ {\bar u}(0) \gamma_\mu \gamma_5 d (0)] | \rho \rangle $.  It is
found that the sum rule for $f_t$ satisfies the in-medium 
Gell-Mann$-$Oakes$-$Renner (GOR) relation precisely while the $f_s$ sum rule 
does not.  The $f_s$ sum rule contains the non-negligible contribution from 
the dimension 5 condensate $\langle {\bar q} i D_0 iD_0 q \rangle_N + 
{1\over 8} \langle {\bar q} g_s \sigma \cdot {\cal G} q \rangle_N$ in addition
to the in-medium quark condensate.  Using standard set of QCD parameters and 
ignoring the in-medium change of the pion mass, we obtain $f_t =105$ MeV at 
the nuclear saturation density.  The prediction for $f_s$ depends on values of 
the dimension 5 condensate and on the Borel mass.  However, the OPE 
constrains that $f_s/f_t \ge 1 $, which does not agree with the prediction
from the in-medium chiral perturbation theory.  Depending on the value of the 
dimension 5 condensate, $f_s$ at the saturation density is found to be in the 
range $ 112 \sim 134$ MeV at the Borel mass $M^2 \sim 1$ GeV$^2$. 
 
\end{abstract}

\maketitle

\section{INTRODUCTION}
\label{sec:intro}

The pion weak decay in nuclear matter is expected to
be different from its decay in free space. 
In hadronic models, the in-medium shift of the decay constant
occurs through the scalar meson  
exchange with  background nucleons as well as through isobar-hole (or
nucleon-hole)
excitations that  intermediates a pion and the axial-vector current~\cite{kr}.
The decay constant is directly connected to the renormalization of the induced 
pseudoscalar coupling, which therefore affects muon captures in 
nuclei~\cite{kr,akh,kirch1,akh1}.

One peculiar feature of the in-medium decay constant
is its separation into the time and space components in the 
matter~\cite{kr,akh1}.
As we will see below, the separation seems to be a natural consequence when
a nuclear matrix element of the axial-vector
current involving a pion state is considered in defining
pion decay constants.  The two decay constants are scaled differently
with the nuclear density and the induced pseudoscalar
constant is separated correspondingly~\cite{kr}.
Interference between  the two decay components
may be important for explaining the muon capture data in nuclei.
Therefore, it is necessary to determine  the decay constants in
the matter.
Indeed, Kirchbach and Wirzba~\cite{kirch1} calculated the
in-medium decay constants using the in-medium chiral perturbation
theory developed in Ref.\cite{thorsson}.
It is found that, while the time component is quenched moderately,
the space component is quenched somewhat substantially.
At the nuclear saturation density, its value is only 20 \% of the free
decay constant, which seems to be a too much quenching.

Another important motivation for studying the in-medium decay
constants is in connection with the so-called Brown-Rho scenario~\cite{BR}.
In this scenario, the in-medium decay constant is
a scaling parameter that governs the in-medium reduction of hadron masses
in a very simple manner.  The quenching of the decay constant
can be regarded as a signature of a partial restoration of chiral symmetry.
To estimate the degree of the restoration, it is important to
calculate the in-medium decay constant model-independently.
Such a calculation can be done using QCD sum rules~\cite{SVZ,qsr}.

In this work, we perform QCD sum rule calculation of the in-medium
decay constant using the
pseudoscalar-axial vector (PA) correlation function in nuclear matter.
The in-medium chiral perturbation theory~\cite{kirch1} will be
used to represent the phenomenological side of the correlation function.
In this approach, the low-lying intermediate state involves
an in-medium quasi-pion whose pole is well separated from 
higher resonance poles.  
As we will see, the sum rules constructed from this correlator
satisfy in some limits the constraints coming purely from PCAC. 
Therefore, this correlation function may be
relevant for studying pion properties in the matter. 

In the QCD side, we will use the finite density QCD sum rule techniques
developed in Ref.~\cite{griegel1,griegel,griegel2}.
The finite density QCD sum rule has been used for investigating
the self-energies of the nucleon embedded in the
matter~\cite{griegel1,griegel,griegel2}.  
Similar sum rule has been applied to the calculation of vector
meson masses at finite density~\cite{HL} or  
temperature~\cite{HKL}. It can provide a useful
constraint for various effective model predictions~\cite{flk}. 
A crucial element in the construction of these sum rules is the linear 
density approximation, which allows to write 
nuclear matrix elements in terms of nucleon
matrix elements plus their vacuum expectation values.  In-medium sum rules
therefore are valid as long as the density is low.
Following these in-medium frameworks, we construct
sum rules for the PA correlation function.

This paper is organized as follow. In Section~\ref{sec:def},
we discuss a subtlety in the definition of the in-medium decay constants.
The isospin symmetry must be assumed in the usual separation
of the time and space components of the decay constant.
In Section~\ref{sec:pa}, we construct the phenomenological side
of the PA correlator by following the in-medium chiral
perturbation theory. We then present a few constraints that
have to be satisfied in some limits in Section~\ref{sec:constraints}.
The OPE calculation is given in Section~\ref{sec:ope} and
the sum rule analysis for the decay constants is presented in
Section~\ref{sec:pionsum}.

\section{Separation into two in-medium decay constants}
\label{sec:def}

There is a subtlety in the definition of the pion decay constants in nuclear
matter.
In this section, we address this issue by defining 
the in-medium pion decay constant in close analogies with the one in
vacuum.  
To do this, we first recall that in vacuum, 
the pion decay constant $f_\pi$ is defined by the relation~\cite{adler},
\begin{eqnarray}
\langle 0| {\bar d} \gamma^\mu \gamma_5 u |\pi^+(k) \rangle
=if_\pi k^\mu\ ,~ (f_\pi =131~{\rm MeV})\ . 
\label{pcac}
\end{eqnarray}
A straightforward generalization to the nuclear matter
case 
is to write the following nuclear matrix element into the 
form~\cite{kr,kirch1}
\begin{eqnarray}
\langle \rho| {\bar d} \gamma^\mu \gamma_5 u |\pi^+(k) \rho \rangle 
= (if_t k_0, if_s {\bf k})\ ,
\label{general}
\end{eqnarray} 
where $|\rho \rangle$ denotes the ground state of the nuclear matter.
This is a usual definition that can be found in literature
but one can show that taking this definition for the in-medium pion decay
constants is equivalent to 
assuming an exact isospin symmetry in the isospin symmetric matter.
The appearance of the two decay constants, $f_t$ and
$f_s$, is also related to the isospin symmetry.

To illustrate this point in detail, we notice that 
the nuclear matrix element can contain a term
proportional to the four-velocity vector $u_\mu$ (with $u^2=1$) of the nuclear
matter as well as a term proportional to the pion momentum $k_\mu$.  
The part proportional to $u_\mu$ can be further
separated into two parts, one that vanishes in the soft-pion limit and
the other that does not.
Hence, we may write the nuclear matrix element into the form, 
\begin{eqnarray}
\langle \rho| {\bar d} \gamma^\mu \gamma_5 u |\pi^+(k) \rho 
\rangle = ia k^\mu + i b u^\mu (u\cdot k)+ic u^\mu \ ,
\label{sep}
\end{eqnarray} 
with some in-medium (dimensionful) constants $a$, $b$ and $c$.
The third term that does not vanish in the soft-pion limit
is proportional to the isospin breaking in the medium. This  can be easily
demonstrated by applying the soft-pion theorem directly to the nuclear 
matrix element, 
\begin{eqnarray}
\langle \rho| {\bar d} \gamma^\mu \gamma_5 u |\pi^+(k) \rho 
\rangle |_{k_\mu \rightarrow 0} \sim 
\langle \rho| {\bar d} \gamma^\mu d  - {\bar u} \gamma^\mu u |\rho \rangle
\sim c u^\mu\ .
\label{spcac}
\end{eqnarray}
Thus, $c$ in Eq.~(\ref{sep}) is proportional to the isospin
breaking in the medium and it is zero either when the isospin symmetry  
is exact or when the nuclear density is zero. 
Also under the isospin symmetry, there are only two in-medium
constants involved for the matrix element Eq.~(\ref{sep}).
In the nuclear rest frame $u^\mu = (1,{\bf 0})$,
we recover Eq.(\ref{general})  
with the identifications~\footnote{One can expect that $a$ and $b$ 
in Eq.~(\ref{sep}) also contain isospin breaking parts. 
As the isospin symmetry is assumed
throughout this work,  we will neglect such terms as well as the
terms belonging to $c$.  Thus, $a$ and $b$ in this equation
are slightly different from the ones that appear in Eq.~(\ref{sep}) by small
isospin breaking parts.} 
\begin{eqnarray}
a+b =f_t \;; \quad a=f_s\ .
\label{rel}
\end{eqnarray}

When the isospin symmetry is broken in the matter, one may quickly notice 
that the time 
component $f_t$ is not well defined.  In this case, the time component
of Eq.~(\ref{sep}) is $iak^0+ibk^0+ic$ and it is not clear
how to deal with the $c$ term in defining $f_t$.
Thus, $f_t$ defined in Eq.~(\ref{general}) is only for the
isospin symmetric case.
The ambiguity in the isospin breaking case is intrinsic in the matter 
and it may be important when the isospin breaking of the in-medium
decays is studied.  
Somewhat larger breaking can be expected from the kaon channel
as the strange vector current is involved.
This breaking in fact makes $K^+$ and $K^-$ decays
different. In future, it will be interesting to
study how the breaking effect is accommodated systematically in meson decays 
in the matter.  In this work,
we will neglect the isospin breaking and focus on the part that
can be represented by Eq.(\ref{general}). 
The isospin breaking terms in the pion sum rules however
are suppressed by the small quark masses and the density.

\section{pseudoscalar and axial vector correlation function}
\label{sec:pa}

In our QCD sum rule study of the in-medium pion decay constants, we consider
the pseudoscalar-axial vector (PA) correlation function in the nuclear medium,
\begin{eqnarray}
\Pi^\mu \equiv i \int d^4x~ e^{ip\cdot x} \langle \rho| 
T[{\bar d}(x) i \gamma_5 u (x)~ {\bar u}(0) \gamma^\mu \gamma_5 d (0)] 
| \rho \rangle\ , 
\label{pcor}
\end{eqnarray}  
where $|\rho \rangle$ denotes the ground state of the nuclear matter.
Because of the separation in Eq.~(\ref{general}),
there are two independent components for this in-medium correlator,
time and space components.
We will consider the following limit of the two components 
\begin{eqnarray}
\lim_{{\bf p}\rightarrow 0}
{\Pi^0 \over i p_0}  \;; \quad \quad 
\lim_{{\bf p}\rightarrow 0} {\Pi^j \over i p^j}
\label{comb}
\end{eqnarray}
so that the resulting functions in the nuclear rest frame depend on one 
scalar variable $p_0^2$.
In this limit, the OPE calculation can be performed 
at $p^2_0 \rightarrow -large$ unambiguously, which is then straightforwardly
matched with the hadronic representation that will be constructed in 
this section.
We will follow Ref.~\cite{kirch1} in making a phenomenological ansatz of 
this correlation function.

Following the in-medium PCAC~\cite{akh1}, the pseudoscalar
current couples to an in-medium quasi-pion. 
Hence, the low-lying state that intermediates 
the PA correlator is expected to be $|\pi(k) \rho \rangle$. 
Of course, other states containing $\pi'$ and $\pi''$
can also intermediate the correlator but we will treat them as the continuum.
Introducing a coupling strength of the pseudoscalar current to the 
in-medium quasi-pion 
$\langle \rho | {\bar d} i \gamma_5 u | \rho \pi \rangle 
\equiv G_\pi^*$ 
and  using  Eqs.~(\ref{sep}) (with $c =0$),  we obtain 
the phenomenological side of the PA correlator~\cite{kirch1},
\begin{eqnarray}
\Pi^\mu =-iG_\pi^*~ {a p^\mu +b u^\mu (u\cdot p) \over
p^2-{m^*_\pi}^2}\ .
\label{copro}
\end{eqnarray}
Here we have introduced the in-medium pion mass $m^*_\pi$.
As we will take the limits given in Eq.~(\ref{comb}),
a possible ${\bf p}$ dependence in $m^*_\pi$ does not affect our
sum rule.  PCAC tells us that in 
vacuum, $G^*_\pi \rightarrow G_\pi =f_\pi m_\pi^2/2m_q$.
In the matter, we should have $G^*_\pi= f_t {m^*_\pi}^2/2m_q$ so that
it satisfies a set of relations given in Ref.~\cite{kirch1}.
Putting Eq.~(\ref{rel}) into Eq.~(\ref{copro}) and 
evaluating it in the nuclear rest 
frame $u^\mu = (1,{\bf 0})$,
we obtain the phenomenological ansatz for the time and space components 
of the correlator
$\Pi^0$ and $\Pi^j$ ($j=1,2,3$) in terms of $f_t$ and $f_s$ 
[defined in Eq.~(\ref{general})], 
\begin{eqnarray}
\Pi^t_{phen} \equiv \lim_{{\bf p}\rightarrow 0}
{\Pi^0 \over i p_0} = 
- {{m^*_\pi}^2\over 2m_q}  {{f_t}^2 \over
p^2_0-{m^*_\pi}^2 } \;; \quad 
\Pi^s_{phen} \equiv \lim_{{\bf p}\rightarrow 0}
{\Pi^j \over ip^j}= - {{m^*_\pi}^2\over 2m_q} {f_t f_s \over
p^2_0-{m^*_\pi}^2 }\ .
\label{combination}
\end{eqnarray}

\section{Constraints}
\label{sec:constraints}

In order to make a reliable prediction for the decay constants
from the PA correlation function,
we need to make sure that the intermediate state is dominated by the
in-medium quasi-pion state. 
This quasi-pion dominance should be checked before one makes some 
claims on the in-medium properties from this PA correlation function.   
Here, we present two stringent constraints that
have to be satisfied for the decay constants.  These constraints will be
compared with our sum rules later in order to justify our approach.

One constraint  is the well-known Gell-Mann$-$Oakes$-$Renner (GOR) relation 
\begin{eqnarray}
-4 m_q \langle {\bar q} q \rangle_0 = m_\pi^2 f_\pi^2\ ,
\label{gor}
\end{eqnarray}
where the subscript denotes the vacuum expectation value.
This has to be satisfied in the ${\cal O}(m_q)$ order whenever we
take the limit $\rho \rightarrow 0$.
When $\rho =0$, we have $f_t=f_s=f_\pi$ and the decay constant is
defined by Eq.~(\ref{pcac}). 
By taking the divergence of the axial-vector current and using the free
equation of motion, we obtain 
\begin{eqnarray}
2 m_q \langle 0| {\bar d} i \gamma_5 u |\pi^+(k) \rangle
=f_\pi^2 m_\pi^2\ . 
\end{eqnarray}
Corrections to this will be of the order ${\cal O}(m_q^2)$ or higher.
The GOR relation Eq.(\ref{gor}) can be easily obtained from this by 
applying the soft-pion theorem ($k_\mu \rightarrow 0$). 

The other constraint is the in-medium GOR relation which will constrain
$f_t$. Starting from the defining 
equation Eq.~(\ref{general}), we take the divergence of the
axial-vector current again assuming that the quasi-pion is
on its mass-shell with $k^2={m^*_\pi}^2$.  
It gives
\begin{eqnarray}
2 m_q \langle \rho | {\bar d} i \gamma_5 u |\pi^+(k) \rho \rangle
=f_t {m^*_\pi}^2 - (f_s - f_t) {\bf k}^2\ . 
\end{eqnarray}
By taking $k$ slightly off the mass shell,
using the reduction formula to convert the pion state into a field,
and following the similar steps as in deriving the soft-pion 
theorem~\cite{Donoghue:1992dd}, we obtain 
\begin{eqnarray}
-{2m_q i \over f_t} \langle \rho | \left [ {\bar d} i \gamma_5 u,
Q_5^+ \right ] |\rho \rangle =
f_t {m^*_\pi}^2 \ ,
\end{eqnarray}
where $Q_5^+$ is the axial charge corresponding to the charged pion.
Note, the soft-pion limit has eliminated the dependence on $f_s$.
Using the anticommutation relation for quarks, we easily evaluate the
commutator to obtain the in-medium GOR relation
\begin{eqnarray}
-4 m_q \langle {\bar q} q \rangle_\rho = {m_\pi^*}^2 f_t^2\ .
\label{mgor}
\end{eqnarray}
Again, this is an exact relation at the order ${\cal O}(m_q)$.
The same relation has been obtained from in-medium chiral perturbation
theory~\cite{kirch1,thorsson,kw}. Note that in deriving these two 
constraints Eqs.(\ref{gor}) (\ref{mgor}), we need to have
a pion (or quasi-pion) state in the matrix elements. It means, 
the quasi-pion dominance in the PA correlation function can be justified
once the resulting sum rules satisfy these constraints.

\section{The operator product expansion (OPE)}
\label{sec:ope}

The OPE calculation for the pseudoscalar-axial vector 
(PA) correlation function Eq.~(\ref{pcor}) is performed
in this section. For a simplicity, we will ignore possible
contributions from instantons~\cite{shuryak} in this PA correlation
function. 
For technical details, we mostly follow 
Refs.~\cite{griegel,Kim:2001ss}.
Before we proceed to the in-medium OPE calculation, it may worth mentioning
on how the OPE looks like in the twist expansion. The leading term in the 
expansion, the 
scalar part, must take the same form as the vacuum sum rule.
In the vacuum case, we know that the correlation
function must reproduce the GOR relation at the order ${\cal O} (m_q)$.
Note, the phenomenological side Eq.~(\ref{combination}) 
contains quark mass in the denominator. To reproduce the GOR relation,  only 
OPE term
that does not vanish in the limit $m_q \rightarrow 0$ must be the
quark condensate.
All others in the OPE  must be zero in the limit $m_q \rightarrow 0$
because the GOR relation
is an exact relation at the order ${\cal O} (m_q)$.  
Modification to this picture in the medium is driven by the higher 
twist operators, which do not need to be scaled with quark mass but they
must be proportional to the nuclear density.

Another thing to mention is that 
the isospin symmetry in the medium is assumed in this work.  It means that
$\pi^+$ or $\pi^-$ decays are assumed to be the same.
Note, once the isospin breaking is introduced, the time component $f_t$
is not well-defined. 
Anyway, under the isospin symmetry, the OPE must be symmetric under 
$u \leftrightarrow d$ as it should be.

The OPE of the correlator Eq.(\ref{pcor})
can be calculated by using the standard techniques in the fixed-point
gauge~\cite{fort}.
To include the finite quark mass effectively, we express 
the correlator in the momentum space. Introducing an in-medium
quark propagator $S_j(k)$ ($j=u,d$) in the momentum space,  we write
Eq.~(\ref{pcor}) as 
\begin{eqnarray}
\Pi^\mu = -i \int {d^4k\over (2\pi)^4} Tr \left [
iS_d(k) i \gamma_5 iS_u(k-p) \gamma^\mu \gamma_5 \right ]\ .
\label{mgpa}
\end{eqnarray}
The trace is taken over the Dirac and color indices.
The perturbative part can be calculated by using the
free propagator
\begin{eqnarray}
iS_j^{free} (k) = i {\fslash {k} + m_{j} \over k^2 - m^2_{j}}~~~(j=u,d)\ 
\label{free}
\end{eqnarray}
in Eq.~(\ref{mgpa}). 
Using standard techniques of the dimensional regularization and
the Feynman parameterization, we readily compute the perturbative part, 
\begin{eqnarray}
\Pi^\mu_{pert} (p^2) =
-{3 \over 4\pi^2 } i p^\mu 
\int^1_0 du~  m_q  {\rm ln} [-u(1-u)p^2 +m_q^2 ]\ 
\label{pert}
\end{eqnarray}
with $m_q = (m_d+m_u)/2$.

Nonperturbative contributions involving the quark condensate can
be obtained by replacing one propagator by the disconnected 
nonlocal quark condensate,
$iS_{ab}^{\alpha \beta} (k) \rightarrow \int d^4x~ e^{ik\cdot x}  
\langle q^\alpha_a (x) {\bar q}^\beta_b (0) \rangle_\rho$ while
the other propagator remains to be the free quark propagator. 
The subscript $\rho$ indicates the in-medium expectation value.
A Tayler expansion around $x_\mu =0$ leads to~\cite{griegel,elias}, 
\begin{eqnarray}
\int d^4x&~& e^{ik\cdot x} \langle q^\alpha_a (x) {\bar q}^\beta_b (0) 
\rangle_\rho
= -{\delta_{ab} \over 12} (2\pi)^4 \left \{\delta^{\alpha \beta} A(k)
+\gamma_\lambda^{\alpha \beta} B^\lambda (k) \right \} 
\label{discon}
\end{eqnarray}
where
\begin{eqnarray}
A(k)&=&\langle {\bar q} q \rangle_\rho \delta^{(4)} (k)
+\langle {\bar q} D_\mu q \rangle_\rho {\partial \over i\partial k^\mu}
\delta^{(4)} (k) 
+ {1\over 2} \langle {\bar q} D_\mu D_\nu q \rangle_\rho 
{\partial \over i\partial k^\mu}{\partial \over i\partial k^\nu}
\delta^{(4)} (k) + \cdot \cdot \cdot\ ,
\nonumber \\
B^\lambda (k)&=&\langle {\bar q}\gamma^\lambda
 q \rangle_\rho \delta^{(4)} (k) 
+\langle {\bar q}\gamma^\lambda D_\mu q \rangle_\rho 
{\partial \over i\partial p^\mu}
\delta^{(4)} (k) 
\nonumber \\
&&+ {1\over 2} \langle {\bar q} \gamma^\lambda 
D_\mu D_\nu q \rangle_\rho 
{\partial \over i\partial k^\mu}{\partial \over i\partial k^\nu}
\delta^{(4)} (k) + \cdot \cdot \cdot\ . 
\end{eqnarray} 

A few technical remarks are in order.
Note that the dimension of the correlator is even. 
In the phenomenological side Eq.~(\ref{copro}),
one dimension is taken up by $p^\mu$ so the rest has
an odd dimension. It means, the constants $a$ and $b$
in Eq.~(\ref{copro}) pick up nonzero contributions only from odd 
dimensional operators in the OPE. Even dimensional operators
are not symmetric under $u \leftrightarrow d$ 
and constitute the isospin breaking parts of the 
OPE which we ignore in this work.
Then, this observation yields interesting consequences.
When the part $A(k)$ in Eq.~(\ref{discon})
is taken for the disconnected quark propagator,
then only the $\fslash {k}$ part of the free quark propagator Eq.~(\ref{free})
survives through the trace in Eq.~(\ref{mgpa}).
In this case, the term like $\langle {\bar q} D_\mu q \rangle_\rho$
constitutes even dimensional contributions, thus cannot contribute
to our isospin symmetric decomposition of Eq.~(\ref{copro}). 
Similarly when the part $B^\lambda$ in Eq.(\ref{discon})
is taken for the disconnected quark propagator,
then the $m_j$ part of the free quark propagator gives
nonzero contributions. Odd dimensional operators in $B^\lambda$ 
(for example $\langle {\bar q} \gamma^\lambda q \rangle_\rho$)
combine with $m_j$ to form
even dimensional operators and they can not
contribute to Eq.~(\ref{copro}). 
In-medium condensates that contribute to our sum rules
can be decomposed into~\cite{griegel}
\begin{eqnarray}
\langle {\bar q} D_\mu D_\nu q \rangle_\rho  &=&
{4\over 3} \langle {\bar q} (u\cdot D)^2 q \rangle_\rho
\left [ u_\mu u_\nu -{1\over 4} g_{\mu \nu} \right ]
-{1\over 3} \langle {\bar q} D^2 q \rangle_\rho
(u_\mu u_\nu - g_{\mu \nu} )\ ,\\
\langle {\bar q}\gamma_\mu D_\nu q \rangle_\rho &=&
{4\over 3} \langle {\bar q} \fslash {u} u\cdot D q \rangle_\rho 
\left [ u_\mu u_\nu -{1\over 4} g_{\mu \nu} \right ]
-{1\over 3} \langle {\bar q}\fslash {u}q \rangle_\rho
(u_\mu u_\nu - g_{\mu \nu} )\ .
\end{eqnarray}

Another nonperturbative contributions come from the quark-gluon mixing operator.
To form such an operator, we need to take
the quark propagator with one gluon line attached~\cite{qsr}
\begin{eqnarray}
iS_{cd} (k) \rightarrow
{g_s\over 2} G^A_{\alpha \beta} t^A_{cd} {1\over (k^2-m^2_q)^3}
(\fslash {k}+m_q) \gamma^\beta (\fslash {k}+m_q) \gamma^\alpha
(\fslash {k}+m_q)\ ,
\end{eqnarray}
where $G^A_{\alpha \beta}$ is the gluon field tensor,
$t_A$ the SU(3) generators in color space with the normalization
$Tr(t^A t^B) = \delta^{AB} / 2$, $c$ and $d$ color
indices. Taking the gluon tensor and moving it into
the disconnected quark propagator,
one can form a quark-gluon mixed condensate in the medium~\cite{griegel}
\begin{eqnarray}
\langle g_s q^\alpha_c {\bar q}^\beta_d G_{\mu \nu}^A \rangle_\rho 
&=&
-{t^A_{cd}\over 96} \Big \{ 
\langle {\bar q} g_s \sigma \cdot {\cal G} q \rangle_\rho
\left [ \sigma_{\mu \nu} + 
i (u_\mu \gamma_\nu - u_\nu \gamma_\mu) \fslash {u} \right ]^{\alpha \beta}
\nonumber \\
&+& \langle g_s {\bar q} \fslash {u} \sigma \cdot {\cal G} q \rangle_\rho
\left [ \sigma_{\mu \nu} \fslash {u} + 
i (u_\mu \gamma_\nu - u_\nu \gamma_\mu) \right ]^{\alpha \beta}
\nonumber \\
&-&4 \left [ \langle {\bar q} (u\cdot D)^2 q \rangle + 
im_q \langle {\bar q} \fslash {u} u\cdot D q \rangle \right ]
\left [ \sigma_{\mu \nu} + 2i (u_\mu \gamma_\nu - u_\nu \gamma_\mu) 
\fslash {u} \right ]^{\alpha \beta} \Big \} \ .
\end{eqnarray}
Here note, ${\cal G}_{\mu \nu} \equiv G^A_{\mu \nu} t^A$.

Combining all these ingredients, we readily compute nonperturbative 
parts that can contribute up to dimension 5. 
After a lengthy but straightforward calculation, we obtain  
\begin{eqnarray}
\Pi^\mu_{nonpert}
&=&ip^\mu {2 \langle {\bar q} q \rangle_\rho \over p^2-m_q^2}
+
{4\over 3} m_q \langle {\bar q} \fslash{u}  u\cdot D q \rangle_\rho 
{p^\mu - 4 p\cdot u u^\mu \over
(p^2 -m^2_q)^2} \nonumber \\ 
&&- 
{2i\over 3} m_q^2 \langle {\bar q} q \rangle_\rho
\left [ {p^\mu -2p \cdot u u^\mu \over (p^2-m^2_q)^2 } + 
{4 (p\cdot u)^2 p^\mu \over (p^2-m^2_q)^3}
\right ]
\nonumber \\
&&+ \left [ {32i\over 3} \langle {\bar q} (u\cdot D)^2 q \rangle_\rho
- {4i\over 3} 
\langle {\bar q} g_s \sigma \cdot {\cal G} q \rangle_\rho
\right ]
\left [ {p\cdot u u^\mu \over (p^2-m^2_q)^2} -
{(p\cdot u)^2 p^\mu \over (p^2 -m^2_q)^3} 
\right ]
\nonumber \\
&&- {16\over 3} m_q \langle {\bar q} \fslash {u} u\cdot D q \rangle_\rho
{p\cdot u u^\mu - p^\mu/4 \over (p^2-m_q^2)^2}
+ {4\over 3} m_q \langle {\bar q} \fslash {D} q \rangle_\rho 
{ p\cdot u u^\mu -p^\mu \over (p^2-m^2_q)^2}\ .
\label{nonpert}
\end{eqnarray}
Here, we have not written down even dimensional operators belonging to 
the isospin breaking.

At the dimension 5, one can expect another nonperturbative contributions 
(with odd dimensional condensate)
containing $ m_q \left \langle {\alpha_s \over \pi} {\cal G}^2
\right \rangle_\rho$ (or gluonic twist operators times $m_q$). 
To compute this, we use quark propagators with one and two gluon 
attached~\cite{qsr} in the calculation of Eq.~(\ref{mgpa}).
Somewhat lengthy calculations show that the Wilson
coefficient of this term is infrared divergent
when the quark mass goes to zero. 
In the case of the vacuum sum rule,
Ref.~\cite{Bagan:1986zp} proposed  
a simple prescription to extract the gluonic Wilson coefficient.
In our in-medium sum rule, further complications arise due
to gluonic twist operators and the simple prescription
proposed in Ref.~\cite{Bagan:1986zp} needs to be further developed 
for the twist operators.  
However, all these gluonic contributions must contain the quark mass 
(to make dimension 5 condensate)
and they should be small.  They may however be important
in the kaon channel as the strange quark mass is much larger
than $m_u$ or $m_d$, which will be studied in future.

Since we are constructing sum rules for the two scalar functions 
given in Eq.~(\ref{combination}),
we can write down from Eqs.(\ref{pert}) (\ref{nonpert}) the time and space 
components of $\Pi^\mu=\Pi^\mu_{pert} + \Pi^\mu_{nonpert}$
in the nuclear rest frame $u^\mu =(1,{\bf 0})$,
\begin{eqnarray}
\Pi^t_{ope} &\equiv&\lim_{{\bf p}\rightarrow 0} {\Pi^0\over ip_0}
\nonumber \\
&=& -{3 \over 4\pi^2 } \int^1_0 du~ m_q  
{\rm ln} [-u(1-u)p^2_0 +m_q^2 ]
+
{2 \langle {\bar q} q \rangle_\rho \over p^2_0-m_q^2}
+
{8 m_q \langle  q^\dagger i D_0 q \rangle_\rho 
-2 m_q^2 \langle {\bar q} q \rangle_\rho
\over
(p^2_0 -m_q^2)^2} \ , 
\label{time}
\\
\Pi^s_{ope}&\equiv&\lim_{{\bf p}\rightarrow 0} {\Pi^j \over ip^j}
\nonumber \\
&=& -{3 \over 4\pi^2 } \int^1_0 du~ m_q  
{\rm ln} [-u(1-u)p^2_0 +m_q^2 ]
+
{2\langle {\bar q} q \rangle_\rho \over p^2_0-m_q^2}
-
{8\over 3} {m_q \langle q^\dagger iD_0 q \rangle_\rho 
\over
(p^2_0-m^2_q)^2}  - 
{2m_q^2 \langle {\bar q} q \rangle_\rho
 \over (p^2_0-m^2_q)^2 }  
\nonumber \\
&&+ {32 \over 3} \left [  \langle {\bar q} i D_0 iD_0 q \rangle_\rho
+ {1\over 8} 
\langle {\bar q} g_s \sigma \cdot {\cal G} q \rangle_\rho
\right ]
{1 \over (p^2_0 -m^2_q)^2} \ .
\label{space}
\end{eqnarray}
In obtaining this, we have neglected the terms whose dimension in the 
numerator is 7.

\section{The in-medium pion decay constants}
\label{sec:pionsum}

Having calculated the OPE for the PA correlation function,
we can easily construct QCD sum rules for the in-medium  pion decay constants 
$f_t$ and $f_s$.
Taking imaginary part of Eqs. (\ref{combination}), (\ref{time}), (\ref{space})
and matching them within the duality region with a Borel weight $e^{-s/M^2}$,
\begin{eqnarray}
\int^{S_0}_0 ds~ e^{-s/M^2} {1 \over \pi} {\rm Im}  
[\Pi^l_{phen} (s) - \Pi^l_{ope} (s)]=0  ~~~~(l=t,s) \ ,
\end{eqnarray}
we obtain sum rules for the time and space components separately,
\begin{eqnarray}
{{m^*_\pi}^2 \over 2 m_q} {f_t}^2 e^{-{m^*}^2_\pi/M^2}
&=&
{3 m_q \over 4 \pi^2} \int^{S_0}_{4m_q^2} d s e^{-s/M^2}
\sqrt {1 - {4 m_q^2 \over s}} - 
2 \langle {\bar q} q \rangle_\rho e^{-m_q^2/M^2}
\nonumber \\
&+& {8m_q \over M^2} \langle q^\dagger iD_0 q \rangle_\rho 
e^{-m_q^2/M^2}
-{2m_q^2 \over M^2} \langle {\bar q} q \rangle_\rho e^{-m_q^2/M^2}\ ,
\label{pion1} \\
{{m^*_\pi}^2 \over 2 m_q} f_t f_s e^{-{m^*}^2_\pi/M^2}
&=&
{3 m_q \over 4 \pi^2} \int^{S_0}_{4m_q^2} d s e^{-s/M^2}
\sqrt {1 - {4 m_q^2 \over s}} - 
2 \langle {\bar q} q \rangle_\rho e^{-m_q^2/M^2}
\nonumber \\
&-& {8 m_q \over 3 M^2 }
\langle q^\dagger iD_0 q \rangle_\rho e^{-m_q^2/M^2}
-{2m_q^2 \over M^2} \langle {\bar q} q \rangle_\rho e^{-m_q^2/M^2}
\nonumber \\
&+&{32 \over 3 M^2 } \left [ \langle {\bar q} i D_0 iD_0 q \rangle_\rho
+ {1\over 8} 
\langle {\bar q} g_s \sigma \cdot {\cal G} q \rangle_\rho \right ]
e^{-m_q^2/M^2}\ .
\label{pion2}
\end{eqnarray}
The continuum threshold is denoted by $S_0$.  The dependence on this 
parameter is expected to be small
as the perturbative part is proportional to the quark mass.

Various nuclear matrix elements appearing in the OPE can be written in terms
of corresponding nucleon matrix elements (denoted by the 
subscript ``$N$'' below) in the linear density approximation.
In this approximation,
the dimension 4 and 5 condensates can be written~\cite{griegel}
\begin{eqnarray}
&&\langle q^\dagger iD_0 q \rangle_\rho 
= \rho~ \langle q^\dagger iD_0 q \rangle_N + 
{m_q \over 4} \langle {\bar q} q \rangle_\rho \ ,
\label{d4}
\\
&&\langle {\bar q} i D_0 iD_0 q \rangle_\rho
+ {1\over 8} 
\langle {\bar q} g_s \sigma \cdot {\cal G} q \rangle_\rho
= \rho \left [ \langle {\bar q} i D_0 iD_0 q \rangle_N
+ {1\over 8} 
\langle {\bar q} g_s \sigma \cdot {\cal G} q \rangle_N
\right ]
+ {m_q^2 \over 4 } \langle {\bar q} q \rangle_\rho\ .
\label{d5}
\end{eqnarray}
These expressions are slightly different from the ones in Ref.~\cite{griegel}
by the quark-mass dependent terms but whose appearance 
can be easily understood from the equation of motion.
Ref.~\cite{griegel} neglected these small quark mass terms. 
It is interesting to
see that, with these quark mass terms, the Wilson coefficient of the 
condensate $m_q^2 \langle {\bar q} q \rangle_\rho$ is delicately canceled
away.  Anyway, in this separation,
the nucleon matrix elements should be understood as the ones 
in the chiral limit $m_q\rightarrow 0$. 
According to Ref.~\cite{griegel}
\begin{eqnarray}
&&\langle q^\dagger iD_0 q \rangle_N \sim 0.18 ~{\rm GeV}
\ ,\\
&&\langle {\bar q} i D_0 iD_0 q \rangle_N
+ {1\over 8} 
\langle {\bar q} g_s \sigma \cdot {\cal G} q \rangle_N
\sim 0.08-0.3~{\rm GeV}^2\ .
\label{others}
\end{eqnarray}
The value of the first equation comes from the twist-2 quark
distribution function in a nucleon.
The value of the second line, the dimension 5 term, is not well-known.
This is related to the twist-3 nucleon distribution function
whose value however can not be reliably calculated.
The bag model estimate gives 0.08 GeV$^2$~\cite{griegel}
while the other standard estimate yields 0.3 GeV$^2$~\cite{griegel}.
In our later analysis, we will take these two values  as a window of 
this dimension 5 condensate. 
 
{\it But it is important to note that this dimension 5 term can not be 
negative.}  To prove this statement, first note that
$\langle {\bar q} g_s \sigma \cdot {\cal G} q \rangle_N =
2 \langle {\bar q} D^2 q \rangle_N$.  Using further that 
$D^2 = D_0^2 -{\bf D}^2$,
one can rewrite  
\begin{eqnarray}
\langle {\bar q} i D_0 iD_0 q \rangle_N
+ {1\over 8} 
\langle {\bar q} g_s \sigma \cdot {\cal G} q \rangle_N
=
{3 \over 4} \langle {\bar q} i D_0 iD_0 q \rangle_N
+ {1 \over 4} \langle {\bar q} (i{\bf D})^2 q \rangle_N\ .
\end{eqnarray}
Note, $iD_0$ and $i{\bf D}$ are hermitian operators  and
therefore their square
must be positive definite. This means that this dimension 5
condensate, as it contains the positive definite operators,
must have the same sign as the positive quantity
$\langle {\bar q}q \rangle_N$.

Using Eqs.(\ref{d4}), (\ref{d5}), we rewrite
Eqs.(\ref{pion1}), (\ref{pion2}) as
(transferring the quark mass from the LHS to the RHS)
\begin{eqnarray}
{m^*_\pi}^2 {f_t}^2 e^{-{m^*}^2_\pi/M^2}
&=&
{3 m_q^2 \over 2 \pi^2} \int^{S_0}_{4m_q^2} d s e^{-s/M^2}
\sqrt {1 - {4 m_q^2 \over s}} - 
4 m_q \langle {\bar q} q \rangle_\rho e^{-m_q^2/M^2}
\nonumber \\
&+& {16 m_q^2 \over M^2} \rho~ \langle q^\dagger iD_0 q \rangle_N 
e^{-m_q^2/M^2} \ ,
\label{fpion1} \\
{m^*_\pi}^2 f_t f_s e^{-{m^*}^2_\pi/M^2}
&=&
{3 m^2_q \over 2 \pi^2} \int^{S_0}_{4m_q^2} d s e^{-s/M^2}
\sqrt {1 - {4 m_q^2 \over s}} - 
4 m_q \langle {\bar q} q \rangle_\rho e^{-m_q^2/M^2}
\nonumber \\
&-& {16 m_q^2 \over 3 M^2 } \rho~
\langle q^\dagger iD_0 q \rangle_N e^{-m_q^2/M^2}
\nonumber \\
&+&{64 m_q \over 3 M^2 } \rho~ 
\left [ \langle {\bar q} i D_0 iD_0 q \rangle_N
+ {1\over 8} 
\langle {\bar q} g_s \sigma \cdot {\cal G} q \rangle_N \right ]
e^{-m_q^2/M^2}\ .
\label{fpion2}
\end{eqnarray}
Our two sum rules mainly differ by 
the  dimension 5 condensate 
$\langle {\bar q} i D_0 iD_0 q \rangle_N
+ {1\over 8} 
\langle {\bar q} g_s \sigma \cdot {\cal G} q \rangle_N$. 
Other terms that make the two sum rules different
are suppressed by higher orders in small quark mass.

As we have announced, let us see if these sum rules
satisfy the constraints discussed in Sec.~\ref{sec:def}.
First in the limit, $\rho \rightarrow 0$, we have
$f_t=f_s=f_\pi$, $m^*_\pi =m_\pi$.  Then, one can easily see that
the two sum rules become identical to  
the vacuum pion sum rule that appears in Ref.~\cite{Kim:2001ss},
\begin{eqnarray}
f_\pi^2 m_\pi^2 e^{-m_\pi^2/M^2} =
{3m_q^2 \over 2\pi^2} \int^{S_0}_{4m_q^2} 
ds~ e^{-s/M^2} \sqrt{1-{4m^2_q \over s}}  
- 4m_q \langle{\bar q} q \rangle e^{-m_q^2/M^2} \ .
\label{vacuumsum}
\end{eqnarray}
This give a mathematical check for our OPE expression.
Furthermore, at the order ${\cal O}(m_q)$, one can see that
the GOR relation is precisely satisfied (Note,
the perturbative part is an order ${\cal O}(m_q^2)$.).

The additional check can be made by observing that the $f_t$ sum rule
satisfies the in-medium GOR relation. Namely,
in the sum rule for $f_t$ Eq.~(\ref{fpion1}), dominant piece 
of the OPE 
comes from $m_q \langle {\bar q} q \rangle_\rho$.  The other
OPE  are down by higher orders in quark mass.
That is, at the order ${\cal O}(m_q)$ 
[or at the order in ${\cal O}({m_\pi^*}^2$)], this sum rule 
precisely reproduces
the in-medium GOR relation [See also Ref.~\cite{kirch1,thorsson,kw}.], 
\begin{eqnarray}
{m^*_\pi}^2 {f_t}^2  \simeq -4 m_q\langle {\bar q} q \rangle_\rho\ . 
\end{eqnarray} 
Such a simple relation can not be obtained from the space component sum rule  
Eq.(\ref{fpion2}) due to the non-negligible dimension 5 term 
$\langle {\bar q} i D_0 iD_0 q \rangle_N
+ {1\over 8} 
\langle {\bar q} g_s \sigma \cdot {\cal G} q \rangle_N$.
Based on these two checks,
the quasi-pion dominance in the PA correlation function can be justified
because the two constraints require a quasi-pion in the intermediate state.

The ratio of Eqs.(\ref{fpion1}) (\ref{fpion2})
gives us an interesting relation for $f_t$ and $f_s$.
Namely neglecting the terms of order ${\cal O}(m_q^2)$ and 
higher, we find the approximate formula,
\begin{eqnarray}
{f_s \over f_t}
\simeq 1- {16 \rho \over 3 M^2}~ {\langle {\bar q} i D_0 iD_0 q \rangle_N
+ {1\over 8} 
\langle {\bar q} g_s \sigma \cdot {\cal G} q \rangle_N \over
\langle {\bar q}q \rangle_\rho }\ .
\label{ratio}
\end{eqnarray}
When $\rho \rightarrow 0$, the ratio becomes the unity as it should be.
Note, $\langle {\bar q}q \rangle_\rho$ is negative and
the dimension 5 condensate in the numerator is, as we have
emphasized, {\it positive}. 
Therefore, we should have 
\begin{eqnarray}
f_s \ge f_t\ .
\label{ourresult}
\end{eqnarray}
One may suspect that the OPE does not converge
fast enough up to dimension 5 and large contributions from
higher dimensional operators can alter this inequality.
But non-negligible higher dimensional operators (even if they exist)
must be related to
twist operators that scale with the density.  
The scalar (nontwist) operators with higher dimensions
must be proportional to ${\cal O} (m_q^2)$ or higher 
orders because the
scalar part of the sum rule at the order ${\cal O} (m_q)$
must satisfy the GOR relation.
Higher twist operators are suppressed either in
the twist expansion and by the high correlator momentum.
For operators having the same twist but with higher spin indices,
we do not have a systematic way to analyze them but
their contributions are usually smaller than the contribution from lower
spin operators~\cite{HL,HKL} in in-medium sum rules.
Even if there are large corrections from higher dimensional operators (though
it is unlikely), the Borel window from which the physical parameter is 
to be extracted
will be chosen such a way that their
contributions are small.
In this window, their contributions  will be down by higher orders 
of the Borel 
mass ($M^4$, $M^6$ and so on) and their effects affect only the region
around the lower boundary of the window. 
Therefore, it is unlikely that such higher dimensional terms reverse the
inequality.

On the other hand, the in-medium chiral perturbation theory
gives the ratio in terms of the chiral perturbation 
parameters(Note, $f_\pi=131$ MeV in our notation.)
\cite{kirch1},
\begin{eqnarray}
{f_s \over f_t} ={ 1 + {4 c_3 \rho \over f^2_\pi} \over
1 + {4(c_2+c_3) \rho \over f^2_\pi} }
=
1-{{4c_2 \rho \over f^2_\pi} \over 1 + {4(c_2+c_3) \rho \over f^2_\pi} }\ .
\label{cpt}
\end{eqnarray}
With the parameter values, $c_2=0.28$ fm and $c_3=-0.55$ fm given 
in Ref.~\cite{kirch1},
the ratio is 0.28 at the nuclear saturation density $\rho =0.17$ 
fm$^{-3}$, substantially smaller than the unity.
If one compares with the free decay constant, the value seems
to be suppressed too much, only 20 \% of $f_\pi$.
Even if we take into account an error in $f_s$ 
coming from a large uncertainty in the parameter $c_3$~\cite{kirch1}, 
our finding is still inconsistent with this result.
To be comparable with this result, the higher dimensional terms 
that we have not calculated
need to be an order of the leading term in the OPE, which certainly
is not realistic.

We now include the higher order $m_q$ corrections in our analysis
and calculate the decay constants at
the nuclear saturation density.
In the phenomenological side, we have three parameters
to be determined $f_t$, $f_s$ and $m^*_\pi$ but we have
only two sum rules. The best fitting method with a trial function of
the form $a e^{-b/M^2}$ may not reliably determine the exponent $b$.
Further input may be needed. 
For simplicity, we take the in-medium pion mass
to be its free value $m_\pi^* =m_\pi$. This assumption is supported by 
Ref.~\cite{kirch1,park,delorme}
where the mass change in the medium is calculated to be about 10 MeV. 
It should be noted that this assumption does not affect one of our results
Eq.~(\ref{ourresult}) as the pion mass is canceled in the ratio. 

The in-medium quark condensate  in the linear density approximation is 
given by~\cite{griegel,griecon} 
\begin{eqnarray}
&& \langle {\bar q} q \rangle_\rho = \langle {\bar q} q \rangle_0
+ \rho~ {\sigma_N \over 2 m_q}\ . 
\end{eqnarray}
The vacuum expectation value is denoted by the 
subscript ``0''. In our analysis, we will use 
$\langle {\bar q} q \rangle_0= -(225~{\rm MeV})^3$ that corresponds to
the quark mass $m_q \sim 7$ MeV~\cite{gasser,leut}. 
The nucleon sigma term is $\sigma_N \sim 45 $ MeV\cite{gasser1}. 
The values for the other nucleon matrix elements are given 
in Eq.(\ref{others}). 
The continuum threshold $S_0$ will be fixed to the $\pi'$ mass
but the sensitivity  of our result on this choice is very small.

In Figure~\ref{fig1}, the solid lines show
the time component of the decay constant $f_t$
with respect to the Borel mass.
The upper one is the vacuum case and the lower one
is from  Eq.~(\ref{fpion1}) at the nuclear saturation density 
$\rho = 0.17$ fm$^{-3}$.
The dashed lines are obtained from the GOR relation.
$f_t$ from the OPE is only 2 \% larger than the one
from the GOR relation. It is clear from this figure that there
is no dependence on the Borel mass.
At the saturation density, we obtain 
\begin{eqnarray}
f_t=105~{\rm MeV}\ ,
\end{eqnarray}
20 \% lower than its free space value. It is not
so different from the value of  the in-medium
chiral perturbation theory $f_t=101$ MeV~\cite{kirch1}.

Figure~\ref{fig2} shows the Borel curve for the ratio $f_s/ f_t$.
The solid curves are from Eqs.(\ref{fpion1}) (\ref{fpion2}) and
the dashed curves are from our approximate formula Eq.({\ref{ratio}),
which are almost indistinguishable from the full curves.
Therefore, higher $m_q$ corrections are negligible.
In getting the upper two curves, we use  the dimension 5 condensate
$\langle {\bar q} i D_0 iD_0 q \rangle_N
+ {1\over 8} 
\langle {\bar q} g_s \sigma \cdot {\cal G} q \rangle_N =0.3 $ GeV$^2$
while the lower two curves use 0.08 GeV$^2$.
The Borel curves are not so stable with respect to the Borel mass.
Therefore, they are not so useful in extracting the precise value of the
ratio.
Higher dimensional operators that we have not included in our calculation 
may affect the Borel curves  at lower mass region and 
stabilize the curves.  Their contributions however should  become smaller
as we go to higher Borel masses. Anyway, because of
this Borel instability, we are not able to predict a precise value for 
the ratio.
As a rough estimate, at $M^2=1$ GeV$^2$, we find the ratio
$f_s/ f_t = 1.28$ 
from the upper curve and 1.07 from the lower curve.
The variation within $0.8~{\rm GeV}^2 \le M^2 \le 1.2~{\rm GeV}^2$ is 
estimated only 3 \% - 10 \%. 
Using our result of $f_t = 105$ MeV, 
the two extremes give the range 
\begin{eqnarray}
112~{\rm MeV} \le f_s \le 134~{\rm MeV}\ .
\end{eqnarray}
Thus, in this rough estimate, the space component of the pion decay 
constant either
slightly increases or decreases from its vacuum value depending upon the
value of the dimension 5 condensate. Of course a more detailed
analysis after including higher dimensional operators may be need
to determine the precise value of $f_s$.

\section{Summary}
\label{sec:sum}

In this work, we have constructed QCD sum rules for 
the pion decay constant in nuclear matter. 
The pseudoscalar-axial vector correlation function in the
matter has been used for this purpose.
The sum rule for $f_t$ is found to 
satisfy the in-medium GOR relation, which indicates that
the method we used is reasonable for investigating the pion
properties in the matter.
The OPE for the space component contains the non-negligible dimension
5 contribution in addition to the quark condensate.
We have established a firm constraint $f_s \ge f_t$
and it does not agree with the result from the in-medium chiral
perturbation theory.
Neglecting the in-medium shift of the pion mass, we 
have obtained $f_t=105$ MeV. Using this value and  from a rough
estimate, the space component of the decay constant is found to be
in the range
$112~{\rm MeV} \le f_s \le 134~{\rm MeV}$.

\acknowledgments
This work was supported by the Korea Research Foundation Grant
KRF-2001-015-DP0104.
We thank Mannque Rho and Su Houng Lee for useful discussions.

\begin{figure}
\caption{ 
The Borel curves for $f_t$, the solid lines from its sum rule and
the dashed lines from the GOR relation. 
The upper two curves are the vacuum case and
the lower two curves are obtained from
the in-medium sum rule at the nuclear saturation density
$\rho =0.17$ fm$^{-3}$.
}
\label{fig1}

\setlength{\textwidth}{6.1in}   
\setlength{\textheight}{9.in}  
\centerline{%
\vbox to 2.4in{\vss
   \hbox to 3.3in{\includegraphics{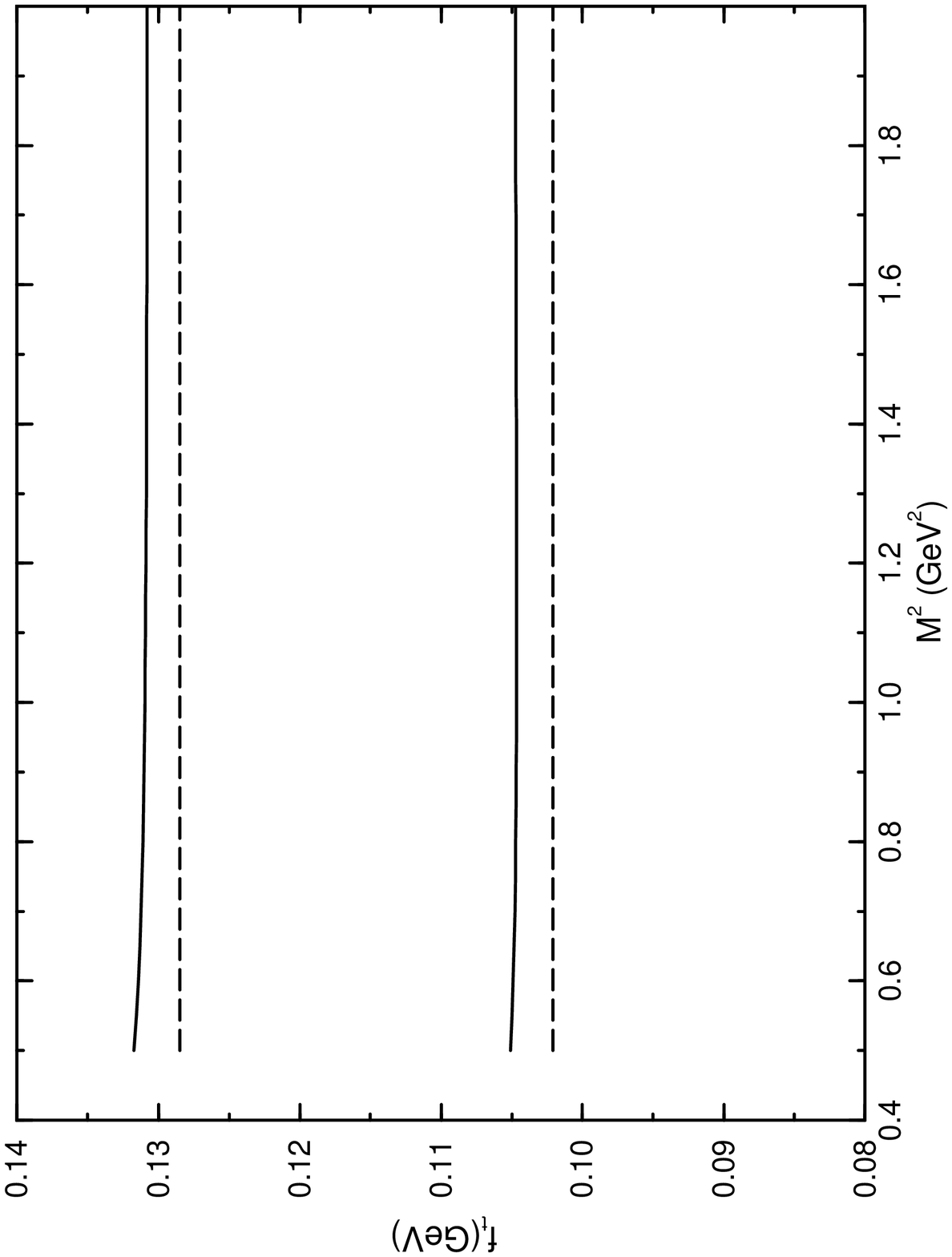}\hss}}
}
\vspace{100pt}
\eject
\end{figure}

\begin{figure}
\caption{The Borel curves for the ratio $f_s/f_t$ at the nuclear saturation
density $\rho =0.17$ fm$^{-3}$.
The upper curves are obtained with  the dimension 5 condensate 0.3 GeV$^2$
and the lower curves are obtained  with 0.08 GeV$^2$.
The solid curves are from our sum rules and the
almost indistinguishable dashed curves are from the
approximate formula Eq.~(\ref{ratio}).
}
\label{fig2}

\setlength{\textwidth}{6.1in}   
\setlength{\textheight}{9.in}  
\centerline{%
\vbox to 2.4in{\vss
   \hbox to 3.3in{\includegraphics{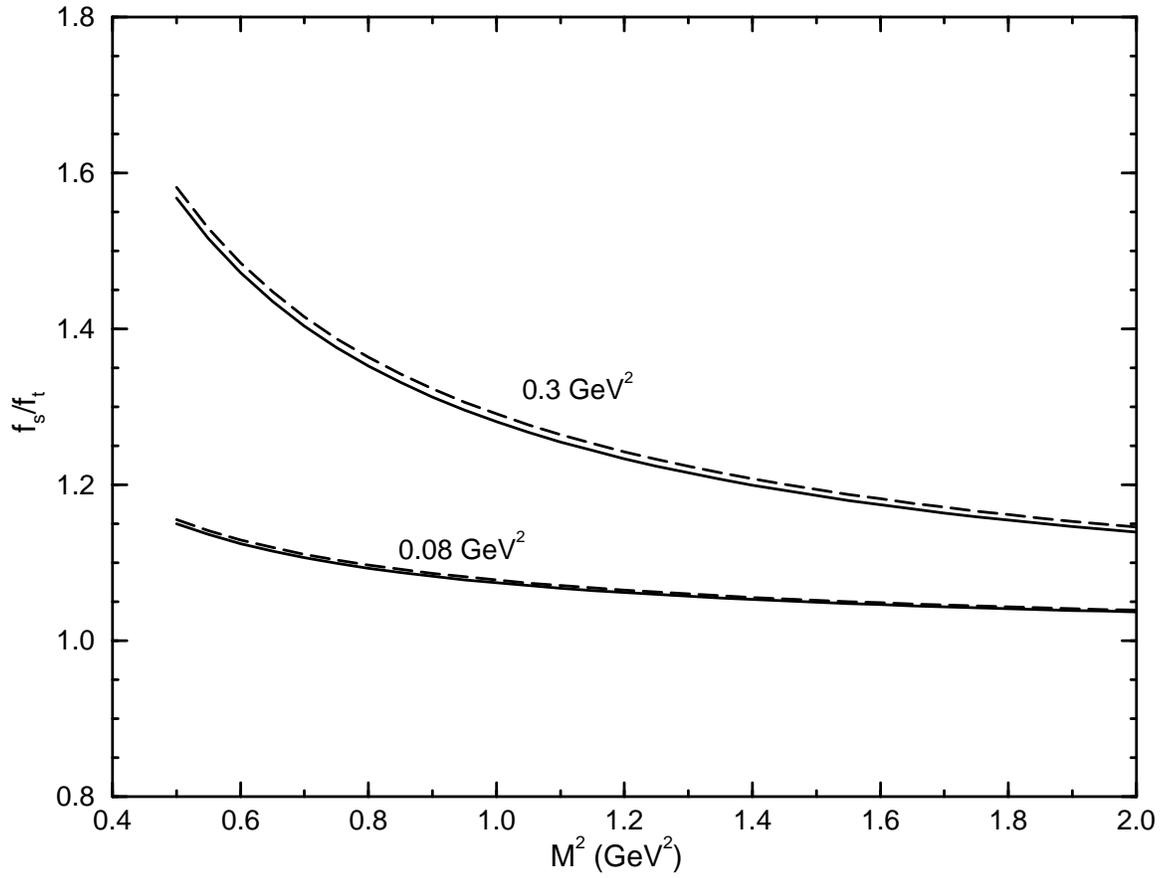}\hss}}
}
\vspace{100pt}
\eject
\end{figure}

\end{document}